\documentclass[twocolumn,preprintnumbers,amsmath,floatfix,prl,aps,superscriptaddress]{revtex4}

\usepackage{graphicx}
\usepackage{epsfig} 
\usepackage{dcolumn}
\usepackage{bm}
\usepackage{times}
\makeatletter

\begin{document} 

\title{Critical Metal Phase at  the Anderson
 Metal-Insulator Transition with Kondo Impurities}

\author{S. Kettemann} 
\affiliation{Jacobs University, School of Engineering and Science,
Campus Ring 1, 28759 Bremen, Germany, and Division of Advanced Materials Science Pohang University of Science and Technology (POSTECH) San31, Hyoja-dong, Nam-gu, Pohang 790-784, South Korea}

\author{E. R. Mucciolo} 
\affiliation{Department of Physics, University of Central Florida,
P. O. Box 162385, Orlando, FL 32816, USA}

\author{I. Varga}
\affiliation{Elm\'eleti Fizika Tansz\'ek, Budapesti M\H uszaki \'es
Gazdas\'agtudom\'anyi Egyetem, H-1521 Budapest, Hungary}

\date{\today}

\begin{abstract}
It is well-known that magnetic impurities can change the symmetry
class of disordered metallic systems by breaking spin and
time-reversal symmetry. At low temperature 
 these symmetries can be restored by Kondo screening.
   It is also known that at the Anderson
metal-insulator transition, wave functions develop multifractal
fluctuations with power law correlations. Here, we consider the
interplay of these two effects. We show that multifractal correlations
open local pseudogaps at the Fermi energy at some random positions in
space. When dilute magnetic impurities are at these locations, Kondo
screening is strongly suppressed. We find that when the exchange
coupling $J$ is smaller than a certain value $J^\ast$, the
metal-insulator transition point extends to a critical region in the
disorder strength parameter and to a band of critical states. The
width of this critical region increases with a power of the
concentration of magnetic impurities.
\end{abstract}



\maketitle

\begin{figure}[t]
\includegraphics[width=8.cm]{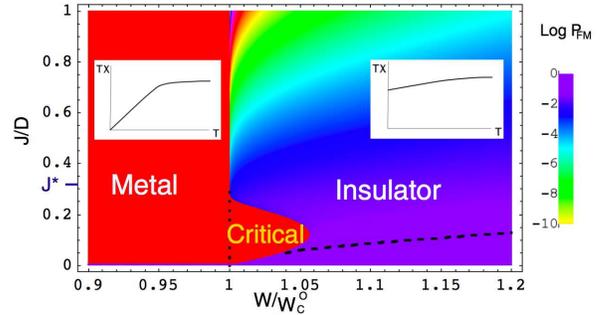}
\caption{(Color online) The fraction of free magnetic moments $P_{\rm
    FM}$ [Eq. (\ref{eq:nFM})] in a three-dimensional disordered electron system
    as function of the exchange coupling $J$ (in units of  band
    width $D$) and the disorder strength $W$ (in units of the critical
    value $W_{\rm c}^{\rm O} $). 
      Dashed line: $J_c^A(W)$,
    Eq. (\ref{jca}),   the   critical exchange coupling caused by local 
    spectral gaps due  to
    Anderson localization.
     For $J < J^\ast \approx 0.3 D$, there is a
    critical metal region for   $W_{\rm c}^{\rm O} < W <
    W_{\rm c} (J)$,  Eq. (\ref{wcj}). 
    Insets: temperature dependence of the local
    magnetic susceptibility $\chi$ of a single magnetic impurity in the metallic 
     and in the insulating phase, showing Kondo screening and free moment behavior, respectively.}
\label{fig:phasediagram}
\end{figure}

  
Fifty years after its proposal, the Anderson metal-insulator
transition (AMIT) of disordered noninteracting electrons
\cite{anderson58} continues to be intensively studied
\cite{brezini,kmck,rmpbelitzkirkpatrick,rmpmirlinevers}. The AMIT is a
quantum phase transition of second order where the localization length
diverges with a critical exponent $\nu$. The critical state is
multifractal and characterized by a wide distribution of wave function
amplitudes with a log-normal shape \cite{rmpmirlinevers}. One aspect
which remains less understood is the interplay between the AMIT of
conduction electrons and the dynamics of the spin of magnetic
impurities. These can be induced by energy levels
below the Fermi energy such as the d-levels of transition metal
impurities  \cite{andersonmm} or by localized electronic states such as the dopant levels in semiconductors \cite{milovanovic}. Magnetic
moments can enhance the spin susceptibility and the specific heat as
has been observed in Si:P close to the AMIT \cite{paalanen,dcmreview}. When
local magnetic moments interact with the conduction electrons by an
exchange interaction, they can break both spin degeneracy and
time-reversal symmetry (TRS) of the conduction electrons. Therefore,
they are expected to change the symmetry class of the electronic
system from orthogonal to unitary \cite{hikami}, changing the critical
exponent $\nu$, the critical electron density $n_c$, and the critical
disorder strength $W_c$ at which the transition occurs \cite{larkin,ohtsukislevinreview}.
At sufficiently low temperatures, an additional effect comes into
play. The antiferromagnetic exchange interaction between spin-$1/2$
local moments and the conduction electrons in a metal leads to a
correlated electron state where a Kondo singlet is formed, 
screening the local moments at temperatures below the Kondo
temperature $T_K$. In this limit, the magnetic susceptibility $\chi$
approaches a constant value, as indicated in the left inset of
Fig. \ref{fig:phasediagram}, and the magnetic moments cease to break
the TRS of the conduction electrons. The
situation is quite different in the insulating phase, where local
spectral gaps $\Delta_I$ prevent the development of the Kondo
screening whenever the exchange coupling $J$ between local moments and
conduction electrons is below a critical value $J_c^A$
\cite{meraikh}. In this case, the unscreened, free magnetic moments
(FMMs) yield a low-temperature susceptibility $\chi \sim 1/T $, as
indicated in the right inset of Fig. \ref{fig:phasediagram}. As a
result, TRS may be partially broken in
the insulating phase. The symmetry class at the AMIT
is determined by the subtle interplay between the Kondo screening and
the Anderson localization.

In this Letter, we aim at determining the density of FMMs in the
vicinity of the AMIT in the low-temperature limit $T \ll \langle T_K
\rangle $ for a system of dilute magnetic impurities with an average
Kondo temperature $\langle T_K \rangle$. An essential aspect of the
problem are the multifractal fluctuations of the eigenfunctions at the
AMIT. They lead to the local suppression of wave function amplitudes
at random locations \cite{multifrac}. Multifractal states are
power-law correlated over a large energy interval $E_c$ of the order
of the band width $D$ \cite{multifrac2,cuevas}. We show that these
spectral correlations induce local pseudogaps which, in turn, prevent
the Kondo screening of a sizeable fraction of local moments even when
the exchange coupling {\it exceeds} $J_c^A$ (see
Fig. \ref{fig:phasediagram}). When the exchange coupling $J$ is
sufficiently weak, these FMMs can break TRS and shift the transition
to stronger disorder amplitudes, $W_c(J)$, as well as lower the critical
exponents on the insulating side to their unitary values. 
Coming from  the metallic side, the transition occurs at the smaller disorder strength $W_c^O$  since  the magnetic impurities become Kondo screened. We conclude that the critical point
is extended to a critical region, as shown in
Fig. \ref{fig:phasediagram}.

The most general model for the formation of magnetic impurities is the
Anderson model \cite{andersonmm} describing a localized level with
energy $\epsilon_d$ and on-site Coulomb repulsion $U$ hybridizing with
electrons in the conduction band of a Hamiltonian $H_0$ which may
include the random potential caused by nonmagnetic impurities $ V({\bf
r})$. Using as basis the eigenstates $\psi_n$ with energy $E_n$ of
$H_0$ and the density operator $\hat{n}_{n,\sigma}$, the Anderson
Hamiltonian can be written as
\begin{eqnarray}
H_A &=& \sum_{n, \sigma} E_n\, \hat{n}_{n,\sigma} + \epsilon_d
\sum_{\sigma} \hat{n}_{d,\sigma} + U\, \hat{n}_{d,+} \hat{n}_{d,-}
\nonumber \\ & & +\ \sum_{n,\sigma} \left( t_{nd}\, c_{n \sigma}^+
d_{\sigma} + \mbox{H.c.} \right).
\end{eqnarray}
Here, $ \hat{n}_{d,\sigma} $ is the density operator in the impurity
level. The hybridization amplitude $t_{nd}$ is proportional to the
eigenfunction amplitude at the position of the magnetic impurity
$\psi_n^* (0)$ and the localized orbital amplitude $\phi_d (0)$:
$t_{nd} = t\, \psi_n^* (0) \phi_d (0)$. One can employ a
Schrieffer-Wolff transformation \cite{hewson,pseudogap} formulated in
terms of eigenstates $\psi_n$ \cite{kettemann06} to generate an s-d
Hamiltonian with exchange couplings
\begin{equation} \label{jnn}
J_{n n^\prime} = t_{n d}^\ast\, t_{n^\prime d} \left( \frac{1}{U +
\epsilon_d - E_{n^\prime}} + \frac{1}{E_n - \epsilon_d} \right)
\end{equation}
and an additional potential scattering term with amplitude
\begin{equation} \label{knn}
K_{n n^\prime} = t_{n d}^\ast t_{n^\prime d} \left( - \frac{1}{U +
\epsilon_d - E_{n^\prime}} + \frac{1}{E_n - \epsilon_d} \right).
\end{equation}
Note that $K_{n n'}$ vanishes for all $n,n^\prime$ when
$|E_{n,n^\prime} - E_F| \ll U$ and $ \epsilon_d = E_F- U/2$, where
$E_F$ is the Fermi energy. For arbitrary $\epsilon_d$ wave functions
with small  amplitude at the position of the magnetic
impurity are hardly modified by this potential scattering term since
$K_{n n^\prime} \sim \psi_n^* (0) \psi_{n^\prime}(0) $ and we
can disregard it when deriving the density of FMMs. Thus, we
retain only the exchange couplings  $J_{n
n^\prime} = J \psi_n^\ast (0) \psi_{n^\prime}(0)$.

{\it Critical exchange coupling.}~-- Recently, it was shown that the distribution 
 of the exchange couplings $J_{n  n\prime}$  results in  a wide bimodal distribution
  of the Kondo temperature and free magnetic moments
\cite{dcmreview,kettemann06,grempel,zhuravlev}. This  was at first  obtained
 by solving the 1-loop Nagaoka-Suhl equation
(NSE) \cite{suhl} of the  Kondo model in the representation
of the eigenstates of $H_0$. It was  confirmed with
the numerical renormalization group and the continuous time quantum
Monte Carlo method \cite{zhuravlev}. These nonperturbative methods
show that the NSE underestimates the density of FMMs. Building on that
work, we  can get the lower limit of the critical
exchange coupling $J_c$ by setting $T_K \rightarrow 0$ in the NSE:
\begin{equation}
\frac{1}{J_c} = \frac{1}{ 2 N} \sum_{n=1}^N \frac{L^d\, |\psi_n ({ 0
}) |^2}{|E_n-E_F|},
\label{jc}
\end{equation}
where $L$ denotes the linear size of the system, $N\sim L^d$ the
number of states in the band, and $d$ is the spatial dimension. For a
clean metallic system with a flat band, the critical exchange coupling
vanishes logarithmically with the number of states $N$ as $J_c \sim
D/\ln N$, leaving no FMMs for any finite value of $J$. On the other
hand, if the eigenstates at the Fermi energy become localized with a
localization length $\xi$, finite local gaps of order $\Delta_I =
(\rho\, \xi^d)^{-1}$ appear at the Fermi energy, cutting off the Kondo
renormalization \cite{meraikh} ($\rho$ is the average
density of states). Thus, there are  FMMs whenever
$\Delta_I \gg T_K$, or, equivalently, $J \ll J_c^{\rm A} \sim D/\ln
N_I$, where $N_{\rm I}= D/\Delta_{\rm I}$ is the number of localized
states with a finite wave function amplitude at the magnetic impurity
site \cite{zhuravlev}.

{\it Critical exchange coupling at the AMIT.}~-- For $d>2$ an AMIT
exists and the localization length diverges at the mobility edge
$E_{\rm ME}$ as a power law, $\xi(E) \sim (|E - E_{\rm
ME}|/E_c)^{-\nu}$. Another way to probe the transition region is to
vary the disorder strength $W$. At a critical strength $W_c$, the
localization length diverges with the same exponent $\nu$. Thus, one
can draw a quantum phase diagram with $J$ versus $W$ where the FMM
phase is limited by a line defined by the function
\begin{equation}
\label{jca}
J_c^A(W)/D = \left[ \nu\, d\, \ln (E_c/|W - W_c|) \right]^{-1},
\end{equation}
as shown by the dashed line in Fig. \ref{fig:phasediagram}.

{\it Local Pseudo Gaps.~--} Equation (\ref{jca}) does not take into
account the multifractality and the critical correlations between wave
functions at different energies at the AMIT \cite{multifrac2}. The
amplitude of multifractal states is suppressed at some random
positions below their typical value, scaling as $L^{-\alpha}$ with
$\alpha > d$ (i.e., decaying faster than extended
states). Correlations between wave functions at different energies can
then open wide local pseudogaps. These correlations can be quantified
by spatially integrating the correlation function of eigenfunction
probabilities associated with two energy levels distant in energy by
$\omega_{nm} =E_n- E_m$, namely \cite{cuevas},
\begin{eqnarray}
\label{criticalcorrelations}
C_{nm} &=& L^d \int d^dr\, \left\langle |\psi_n({\bf r})|^2
|\psi_m({\bf r})|^2 \right\rangle \nonumber \\ &=& \left\{
\begin{array}{ll}
(E_c/\Delta)^\gamma, & |\omega_{nm}| < \Delta, \\
(E_c/|\omega_{nm}|)^\gamma, & \Delta < |\omega_{nm}| < E_c, \\
(E_c/|\omega_{nm}|)^{2}, & |\omega_{nm}| > E_c,
\end{array}
\right.
\end{eqnarray}
where $0<\gamma <1$. For $\omega_{nm} < E_c$ the value of $C_{nm}$ is
enhanced in comparison to the plane-wave limit  $C_{nm}=1$. The
distribution function of a multifractal wave function is radically
different from the Porter-Thomas distribution of metallic states since
the moments of eigenfunction intensities $|\psi_l({\bf r})|^2$ obey
the power-law scaling $P_q = L^d \langle\, |\psi_l({\bf r})|^{2 q}
\rangle \sim L^{-\tau_q}$, where $\tau_q$ is the multifractal exponent
of the $q$-th moment. The corresponding distribution function is known
to be  log-normal, in good approximation \cite{rmpmirlinevers},
\begin{equation}
\label{eq:Pone}
P(\alpha) = L^{ - \frac{(\alpha-\alpha_0)^2}{2 d \gamma }},
\end{equation}
where $\alpha = - \ln |\psi_l({\bf r})|^2 / \ln L$.  $\alpha_0 > d$ is
related to the multifractal exponents, $\tau_q = d (q-1) +
(\alpha_0-d) q (1-q)$, and to $\gamma = 2(\alpha_0-d)/d$. In order to
find $J_c$ we need the joint distribution of two eigenfunction
intensities $|\psi_l({\bf r})|^2$ and $|\psi_n({\bf r})|^2$, $l\neq n$
which is also log-normal,
\begin{equation}
\label{eq:Ptwo}
P(\alpha_l,\alpha_k) = L^{- 2 d + a_{l k} \left[ f (\alpha_l) +
f(\alpha_k) \right] + b_{l k} \frac{(\alpha_l - \alpha_0)(\alpha_k
-\alpha_0) }{d \gamma}},
\end{equation}
with $f(\alpha) = d - (\alpha - \alpha_0)^2/2 d \gamma $. Imposing
that Eq. (\ref{eq:Ptwo}) is consistent with
Eqs. (\ref{criticalcorrelations}), we determine the coefficients as
$a_{k l} = ( 1 + \sqrt{1 + 4 b_{k l}^2} )/2$ and $b_{k l} = g_{k
l}/(g_{k l}^2-1)$, where
\begin{equation}
g_{k l} = \frac{ \ln \left( |\omega_{lk}|/E_c \right)} {d \ln L} \times
\left\{ \begin{array}{lr} 1, & |\omega_{lk}| < E_c, \\ 2/\gamma, & |\omega_{lk}| > E_c.
\end{array} \right.
\end{equation}
The conditional probability to have
$\alpha_l$ when $\alpha_n = \alpha$ is then
\begin{equation} \label{conditional}
P_{\alpha_n=\alpha} (\alpha_l) = \frac{P(\alpha,\alpha_l)}{P(\alpha)}
= L^{- \frac{[\alpha_l - \alpha_0 + g_{ln}(\alpha -
\alpha_0)]^2}{2 d \gamma  \left( 1- g_{ln}^2 \right)}}.
\end{equation}
Averaging Eq. (\ref{jc}) with the conditional probability
Eq. (\ref{conditional}), we find that, at a site where $\alpha_n =
\alpha$ with $E_n \approx E_F$, the critical exchange coupling $J_c$ is
determined by the equation
\begin{equation}
\frac{1}{J_c} = \sum_{E_l = E_F - E_c}^{E_F+E_c} \frac{ ({| E_l -E_F|
/E_c})^{\frac{\alpha -\alpha_0}{d}} L^{-\frac{d \gamma}{2} g_{l n}^2} }{N
|E_l -E_F|}.
\end{equation}
The factor $({| E_l -E_F| /E_c})^{\frac{\alpha -\alpha_0}{d}}$ corresponds to a  local pseudogap with
exponent $(\alpha -\alpha_0)/d$ at positions where $\alpha > \alpha_0$. At such sites the 
critical exchange coupling is  proportional to that exponent,  in the continuum limit $\Delta \sum_l \rightarrow \int
dE_l$,
\begin{eqnarray}
\label{jcL} 
\frac{J_c (\alpha)}{D} & = & \frac{\alpha - \alpha_0}{d} \left\{ 1-
\exp \left[ - \frac{(\alpha - \alpha_0) \sqrt{\ln L}}{ \sqrt{2 d \gamma}} \right] \right\}^{-1}.
\end{eqnarray}
Notice that when $\alpha$ takes its typical value $\alpha_0$, we get
$J_c(\alpha_0) =  \sqrt{2 d \gamma/d \ln L}$, which decreases
slowly with increasing $L$. At sites where $ \alpha \le \alpha_0$, the
LDOS is larger than its typical value and $J_c$ vanishes. The exchange
coupling $J_c^{(1)}$ at which there is on average only one FMM in the
whole system is $J_c^{(1)} = J_{c}(\alpha_+) = \sqrt{2 \gamma} D$,
where $\alpha_+$ is obtained from the condition $P(\alpha_+) = 1/L^d$,
yielding $\alpha_+ = \alpha_0 + \sqrt{2 \gamma } d$. Similarly,
at $J_c^{(2)} = \sqrt{\gamma} D$, one finds $O(\sqrt{N})$ sites with
FMMs. A finite density of FMMs at the AMIT is only found at very small
exchange couplings, $J<J_c(\alpha_0)$, and these values are
vanishingly small for $L \rightarrow \infty$.
 
{\it Pseudogaps and FMMs in the insulating phase.}~--While we conclude
that there is no finite density of FMMs at the AMIT in the
infinite-volume limit, the situation changes on the insulating side of
the transition. First,  a hard spectral gap of
order $\Delta_I$ quenches all magnetic moments below an exchange
coupling $J_c^A$, given by Eq. (\ref{jca}). However, there are still
power-law spectral correlations between wave functions located within
the same localization volume $\xi^d$. One can take these correlations
into account by noting that the wave function intensities within a
localization volume still have a log-normal distribution  with $\alpha \rightarrow \alpha_{\xi} =
- \ln |\psi|^2/\ln \xi$. For the evaluation of
$J_c$ we therefore should substitute $L$ by the localization length $\xi(W) \sim
(W- W_c)^{- \nu}$. Thus, for fixed $J$, the density of FMMs 
 depends on the localisation length $\xi$ as
\begin{equation}
\label{eq:nFM}
P_{\rm FM} = n_{\rm FM} /n_M = {\rm Erfc} \left( \sqrt{\frac{\ln
\xi}{2 \gamma}}  \frac{J}{D} \right).
\end{equation}
Close to the transition,  $\xi$ is large and
Eq. (\ref{eq:nFM}) simplifies to $P_{\rm FM} \sim (W -W_c)^{
\kappa(J)}$, decaying with an exponent $\kappa(J) = (\nu d/2 \gamma) (
J/D)^2$.

    
{\it Metallic phase.}~-- On the metallic side of the AMIT, the
correlation length $\xi_c$ limits the range of multifractal
correlations. One can imagine the extended states close to the
transition as patches of multifractal states connected to each other
by tunneling \cite{efetov}. This yields the scaling $|\psi|^{2 q} \sim
(\xi_c/L)^{q d} \xi_c^{-d -d_q} \sim L^{-q d}$ for $ L > \xi_c$,
leaving no finite density of FMMs in the metallic phase ($W< W_c$).

{\it Symmetry Class of the AMIT.}~-- The magnetic scattering by FMMs
changes the symmetry class of the conduction electrons from orthogonal
to unitary when $\xi^2/{\cal D} \tau_s >1$, where ${\cal D}$ is the
diffusion constant, $1/\tau_s$ is the magnetic scattering rate, and
$\xi$ denotes  the correlation length on the metallic side, or the 
localization length on the insulating side of the transition, respectively
\cite{hikami,larkin,droese}. The magnetic scattering rate at zero
temperature is bounded from below by $ 2\pi n_{FM} S(S+1) (J/D)^2 \rho
(\epsilon_F, {\bf r})/\rho^2$. Coming from the metallic side of the
transition, $n_{FM}$ vanishes due to the Kondo screening and the AMIT
occurs at the orthogonal critical value for time-reversal symmetric
systems, $W_{\rm c}^{\rm O}$.
%
%
%
On the insulating side of the AMIT, the density of FMMs is finite, as
given by Eq. (\ref{eq:nFM}). The LDOS at the position of the FMMs
scales as $ \rho (\epsilon_F, {\bf r}) \sim \xi^{d-\alpha(J)}$, where,
$\alpha(J)= \alpha_0+ d J/D$. Therefore, $1/\tau_s$ is finite and,
following Refs. \cite{larkin,droese}, we can get the critical disorder
amplitude $W_c(J) $  from the condition $D
\tau_s(J) = \xi \left(W_c(J) -W_{\rm c}^{\rm O}\right)^2$:
\begin{equation}
\label{wcj}
\frac{W_{\rm c} (J)- W_{\rm c}^{\rm O}}{W_{\rm c}^{\rm O}} = \left[
\frac{2^d (\pi W_c(J) J)^2\, n_M\, S(S+1)}{ d D^4} \right]^{\eta(J)} ,
\end{equation}
where $1/\eta(J) = 2 \nu - (d \nu/2 \gamma) (\gamma + J/D)^2$. For $J$ exceeding $J^\ast = (2
\sqrt{\gamma/d}-\gamma) D$,  the critical disorder amplitude $W_{\rm
c}(J)$ approaches its orthogonal value $W_{\rm c}^{\rm O}$. 
However, for
smaller exchange couplings, $J<J^\ast$, a paradoxical situation appears:

{\it The Critical Metal Phase.}~--
 The position of the critical point
$W_{\rm c}$ depends on the direction from which the AMIT is
approached: Coming from the metallic side, the Kondo effect screens
all magnetic moments at low temperature $T \ll T_K$, so that TRS is
not broken and the AMIT occurs at $W_{\rm c}^{\rm O}$. Coming from the
insulating side, there is a finite density of FMMs which break TRS and
lead to an AMIT at the stronger disorder
amplitude $W_{\rm c} (J)$ given by Eq. (\ref{wcj}). Consistency requires then that these  two distinguished 
critical points are
connected  for intermediate disorder strengths $W_{\rm
c}^{\rm O}< W < W_{\rm c} (J)$  by a {\it critical metal region}.
 Accordingly, the
mobility edge is extended to a {\it critical band} whose width is a
function of $J$ and $n_M$. The resulting zero-temperature quantum
phase diagram is shown in Fig. \ref{fig:phasediagram} with the
critical metal phase appearing below the tricritical point $J^\ast$.
The analytical expression for the density of FMMs compares
quantitatively well with results obtained from numerical simulations
of critical random banded matrices \cite{varga08}, and are in
qualitative agreement with numerical simulations of the
three-dimensional AMIT \cite{grempel}. Furthermore, we have derived \cite{varga08}
the distribution of the Kondo temperature analytically taking into account 
multifractality and critical correlations of the critical wave
functions and found a wide bimodal distribution at the AMIT.
 In two dimensions, the distribution
is also bimodal with a finite width which increases with the second
power of the disorder amplitude, in qualitative agreement with
numerical simulations \cite{grempel,zhuravlev}.

A wide distribution of Kondo temperatures was argued previously to cause a finite
electron dephasing rate at low temperatures \cite{kettemann06} and to
lead to deviations from the Fermi-liquid behavior 
\cite{dcmreview,milovanovic}. Here we consider the zero-temperature
limit and find that on the metallic side of the AMIT the density of FMMs vanishes due to the Kondo
screening. However, a finite density of FMMs survives on the insulating side of the AMIT. 
 A new critical metal phase arises due to  the
interplay between the Kondo effect and the multifractality of the wave
functions. One
experimental consequence is the divergence of the dielectric
susceptibility, $\chi (T=0)  \sim \xi^2 \sim
(W_c(J)-W)^{-2\nu}$ at the disorder amplitude $W_c(J)$ (or at a
critical electron density $n_c(J)$), while the zero temperature limit of the resistivity diverges
with the correlation length as $\rho (T=0)  \sim \xi_c \sim (W-W_c^O)^{-\nu}$
at the weaker disorder amplitude $W_c^O$ (or at a larger critical
density $n_c^O$, accordingly). This difference $W_c(J) - W_c^O$
increases with the concentration of Kondo impurities as 
Eq. (\ref{wcj}), maximally by about   $10 \%$\cite{ohtsukislevinreview}. We expect the critical metallic phase to be
observable in materials where both the AMIT and the Kondo
effect are present simultaneously at experimentally accessible temperatures, such as in  amorphous
metal-semiconductor alloys \cite{kmck,amormetal} with dilute magnetic impurities\cite{hasegawa},
 or in doped semiconductors, such as Si:P where thermopower measurments 
 are consistent with  the presence of Kondo impurities with  $ \langle T_K
\rangle \approx 1 K$\cite{thermopower}.

This research was supported by the   German Research Council under SFB
668 B2, the Alexander von Humboldt Foundation, the Hungarian Research
Fund (OTKA) under K73361 and K75529,  the International Center for
Transdisciplinary Studies at Jacobs University, and  the WCU  program
of the Korea Science and Engineering Foundation under
Project No. R31-2008-000-10059-0.



\begin{thebibliography}{11}
\bibitem{anderson58} P. W. Anderson, Phys. Rev. {\bf 109}, 1492
(1958).

\bibitem{brezini} A. Brezini and N. Zekri, Phys. Stat. Sol. B {\bf
169}, 253 (1992).

\bibitem{kmck} B. Kramer and A. MacKinnon, Rep. Prog. Phys. {\bf 56},
1469 (1993).

\bibitem{rmpbelitzkirkpatrick} D. Belitz and T. R. Kirkpatrick,
Rev. Mod. Phys. {\bf 66}, 261 (1994).

\bibitem{rmpmirlinevers} F. Evers and A. D. Mirlin, Rev.
Mod. Phys. {\bf 80}, 1355 (2008).

\bibitem{andersonmm} P. W. Anderson, Phys. Rev. {\bf 124}, 41(1961).

\bibitem{milovanovic} M. Milovanovic, S. Sachdev, and R. N. Bhatt,
Phys. Rev. Lett. {\bf 63}, 82 (1989);

\bibitem{paalanen} M. A. Paalanen and G. A. Thomas, Helv. Phys. Acta
{\bf 56}, 27 (1983); M. A. Paalanen, J. E. Graebner, R. N. Bhatt, and
S. Sachdev, Phys. Rev. Lett. {\bf 61}, 597 (1988); S. Sachdev,
Phys. Rev. B {\bf 39}, 5297 (1989).

\bibitem{dcmreview} E. Miranda and V. Dobrosavljevi\'c,
Rep. Prog. Phys. {\bf 68}, 2337 (2005).

\bibitem{hikami} S. Hikami, A. I. Larkin, and Y. Nagaoka,
Prog. Theor. Phys. {\bf 63}, 707 (1980).

\bibitem{larkin} D. Khmelnitskii and A. I. Larkin, Solid State
Comm. {\bf 39}, 1069 (1981).


\bibitem{ohtsukislevinreview} T. Ohtsuki and T. Kawarabayashi,
J. Phys. Soc. Jpn. {\bf 66}, 314 (1997); T. Ohtsuki, K. Slevin, and
T. Kawarabayashi, Ann. Physik {\bf 8}, 655 (1999).


\bibitem{meraikh} S. Kettemann and M. E. Raikh, Phys. Rev. Lett. {\bf
90}, 146601 (2003).

\bibitem{multifrac} F. Wegner, Z. Phys. B {\bf 36}, 209 (1980);
  H. Aoki, J. Phys. C {\bf 16}, L205 (1983); C. Castellani and
  L. Peliti, J. Phys. A: Math. Gen. {\bf 19}, L991 (1986);
  M. Schreiber and H. Gru\ss bach, Phys. Rev. Lett. {\bf 67}, 607
  (1991); M. Janssen, Int. J. Mod. Phys. B {\bf 8}, 943 (1994).

\bibitem{multifrac2}
J. T. Chalker, Physica {\bf 167A}, 253 (1990);
%
V. E. Kravtsov and K. A. Muttalib,
Phys. Rev. Lett. {\bf 79}, 1913 (1997);
%
J. T. Chalker, V. E. Kravtsov, and I. V. Lerner, JETP
Lett. {\bf 64}, 386 (1996); V. E. Kravtsov, Ann. Phys. (Leipzig) {\bf
8}, 621 (1999).

\bibitem{cuevas} E. Cuevas and V. E. Kravtsov, Phys. Rev. B {\bf 76},
235119 (2007).

\bibitem{pseudogap} D. Withoff and E. Fradkin, Phys. Rev. Lett. 64,
1835 (1990); K. Ingersent, Phys. Rev. B 54, 11936 (1996); L. Fritz,
S. Florens, and M. Vojta, Phys. Rev. B 74, 144410 (2006).

\bibitem{hewson} A. C. Hewson, {\it The Kondo Problem to Heavy
 Fermions}, Cambridge Univ. Press (1997).



\bibitem{kettemann06} S. Kettemann and E. R.  Mucciolo, JETP
Lett. {\bf 83}, 240 (2006) [Pis'ma v ZhETF {\bf 83}, 284 (2006)];
S. Kettemann and E. R.  Mucciolo,
Phys. Rev. B {\bf 75}, 184407 (2007); T. Micklitz, T. A. Costi, and
A. Rosch, {\it ibid}. B {\bf 75}, 054406 (2007),
T. Capron {\it et al.}, {\it ibid}. {\bf 77}, 033102 (2008).

\bibitem{grempel} P. S. Cornaglia, D. R. Grempel, and C. A. Balseiro,
Phys. Rev. Lett. {\bf 96}, 117209 (2006).

\bibitem{zhuravlev} A. Zhuravlev,
 I. Zharekeshev, E. Gorelov, A. I. Lichtenstein, E. R. Mucciolo and S. Kettemann,
Phys. Rev. Lett. {\bf 99}, 247202 (2007).

\bibitem{suhl} Y. Nagaoka, Phys. Rev. {\bf 138}, 1112 (1965); H. Suhl,
Phys. Rev. {\bf A 138}, 515 (1965).

\bibitem{efetov} K. B. Efetov, {\it Supersymmetry of Disorder and
  Chaos} (Cambridge University Press, Cambridge, 1997).



\bibitem{droese} T. Dr\" ose, M. Batsch, I. Kh. Zharekeshev, and
B. Kramer, Phys. Rev. B {\bf 57}, 37 (1998).




\bibitem{varga08} I.~Varga, E.~R.~Mucciolo, and S.~Kettemann,
unpublished (2009).

\bibitem{amormetal} C. Van Haesendonck and Y.  Bruynseraede,
Phys. Rev. B {\bf 33}, 1684 (1986); B. W. Dodson, W. L. McMillan,
J. M. Mochel, and R. C. Dynes, Phys. Rev. Lett. {\bf 46}, 46 (1981).


\bibitem{hasegawa} C. C: Tsuei and R. Hasegawa, Solid State Commun. {\bf 7}, 1581 (1969). 



\bibitem{thermopower}   M. Lakner and H. v. Lohneysen,
Phys. Rev. Lett. 70, 3475 (1993).
\end{thebibliography}
\end{document}